\documentclass[showpacs,preprintnumbers]{revtex4}
\usepackage{amsmath,amssymb}
\usepackage{graphicx}
\usepackage{dcolumn}
\usepackage[vcentermath]{youngtab}
\usepackage[dvips]{epsfig,color}
\usepackage{bm}
\begin{document}
\title{Relativistic quark-diquark model of baryons. Non strange spectrum and nucleon electromagnetic form factors}
\author{M. De Sanctis}
\affiliation{Universidad Nacional de Colombia, Bogot\'a, Colombia} 
\author{J. Ferretti}
\affiliation{INFN, Sezione di Genova, via Dodecaneso 33, 16146 Genova (Italy)}
\affiliation{Instituto de Ciencias Nucleares, Universidad Nacional Aut\'onoma de M\'exico, Distrito Federal, M\'exico}
\author{E. Santopinto}\thanks{Corresponding author: santopinto@ge.infn.it}\author{A. Vassallo}
\affiliation{INFN, Sezione di Genova, via Dodecaneso 33, 16146 Genova (Italy)}   
\begin{abstract}
We briefly describe our relativistic quark-diquark model, developed within the framework of point form dynamics, which is the relativistic extension of the interacting quark-diquark model. In order to do that we have to show the main properties and quantum numbers of the effective degree of freedom of constituent diquark. Our results for the nonstrange baryon spectrum and for the nucleon electromagnetic form factors are discussed. 
\end{abstract}
\maketitle

\section{Introduction}
The concept of diquark was introduced by Gell-Mann in 1964 \cite{GellMann:1964nj} and soon afterwards the effective degrees of freedom of diquarks were used in order to describe baryons as bound states of a constituent quark and diquark \cite{Ida:1966ev}.
There are several phenomenological arguments in favor of diquark correlations \cite{Wilczek:2004im,Jaffe:2004ph,Selem:2006nd}. For example, it is known that the quark-diquark model does not predict the existence of the $\mbox{{\boldmath{$20$}}}$ SU$_{\mbox{sf}}$(6) multiplet, which on the contrary is predicted by three quarks constituent quark models.
If one decomposes such SU(6)$_{\mbox{sf}}$ multiplet in terms of SU$_{\mbox{f}}$(3) $\otimes$ SU$_{\mbox{s}}$(2), one gets a spin-$\frac{1}{2}$ flavor octet and a spin-$\frac{3}{2}$ flavor singlet: the absence of the spin-$\frac{3}{2}$ flavor singlet from PDG \cite{Nakamura:2010zzi} represents an argument in favour of quark-diquark models. 
There are also phenomenological indications, regarding the Regge behavior of hadrons, the $\Delta I = \frac{1}{2}$ rule in weak nonleptonic decays \cite{Neubert}, some regularities in parton distribution functions \cite{Close:1988br} and in spin-dependent structure functions \cite{Close:1988br} and in $\Lambda(1116)$ and $\Lambda(1520)$ fragmentation functions \cite{Wilczek:2004im,Jaffe:2004ph,Selem:2006nd}.
Moreover it is well known that any interaction, that is strong enough to bind $\pi$ and $\rho$ mesons in the so called rainbow-ladder approximation of QCD's Dyson-Schwinger Equation (DSE), will also produce diquarks \cite{Cahill:1987qr}. Furthermore indications for diquark confinement have also been provided \cite{Bender:1996bb}. This makes plausibly enough to make diquarks a part of the baryon's wave function. 

Other interesting approaches to baryon (meson) spectroscopy and structure, studied within other types of models and/or formalisms, can be found in Refs. \cite{other-refs}.

\section{Nonrelativistic quark-diquark states}
We assume that baryons are composed of a constituent diquark and quark.
The effective degree of freedom of diquark describes two strongly correlated quarks, without internal spatial excitations (thus in $S$ wave). 
The diquark has to be in the $\mbox{{\boldmath{$\bar{3}$}}}$ representation of SU$_{\mbox{c}}$(3), since the baryon must be colorless. Thus the possible SU$_{\mbox{sf}}$(6) representations for the diquark are limited to the symmetric $\mbox{{\boldmath{$21$}}}$ representation, that contains the $[\mbox{{\boldmath{$\bar{3}$}}},0]$ and $[\mbox{{\boldmath{$6$}}},1]$ representations. The notation used here is $[\mbox{{\boldmath{$F$}}}_1, s_1]$, where $\mbox{{\boldmath{$F$}}}_1$ stands for the dimension of the SU$_{\mbox{f}}$(3) representation and $s_1$ for the spin of the diquark.
If one calculates the quark-quark interaction due to one gluon exchange, the channel $\mbox{{\boldmath{$\bar{3}$}}}_f$ looks energetically favoured \cite{Wilczek:2004im,Jaffe:2004ph,DeRujula:1975ge,DeGrand:1975cf} and thus such configuration, i.e. $[\mbox{{\boldmath{$\bar{3}$}}},0]$ (spin singlet, antisymmetric in flavor), is defined by Wilczek \cite{Wilczek:2004im} and Jaffe \cite{Jaffe:2004ph} as "good" diquark (or scalar diquark), while the other one, i.e. $[\mbox{{\boldmath{$6$}}},1]$ (spin triplet, symmetric in flavor), is defined as "bad" diquark (or axial-vector diquark).

If only nonstrange baryons are considered, one has the following basis in spin-isospin space,
\begin{subequations}
\label{eqn:NR-Spin-isospin-states}
\begin{equation}
	\begin{array}{c}
	\left|~s_1~=~0,~t_1~=~0;~\frac{1}{2}~,~\frac{1}{2};~S~=~\frac{1}{2}, ~T~=~\frac{1}{2}~\right\rangle~~~,  
	\end{array}
\end{equation}
\begin{equation}
	\begin{array}{c}
	\left|~s_1~=~1,~t_1~=~1;~\frac{1}{2},~\frac{1}{2};~S~=~\frac{1}{2}, ~T~=~\frac{1}{2}~\right\rangle~~~,  
	\end{array}
\end{equation}	
\begin{equation}	
	\begin{array}{c}
	\left|~s_1~=~1,~t_1~=~1;~\frac{1}{2},~\frac{1}{2};~S~=~\frac{1}{2}, ~T~=~\frac{3}{2}~\right\rangle~~~,  
	\end{array}
\end{equation}	
\begin{equation}	
	\begin{array}{c}
	\left|~s_1~=~1,~t_1~=~1;~\frac{1}{2},~\frac{1}{2};~S~=~\frac{3}{2}, ~T~=~\frac{1}{2}~\right\rangle~~~,  
	\end{array}
\end{equation}	
\begin{equation}	
	\begin{array}{c}
	\left|~s_1~=~1,~t_1~=~1;~\frac{1}{2},~\frac{1}{2};~S~=~\frac{3}{2}, ~T~=~\frac{3}{2}~\right\rangle~~~,
	\end{array}
\end{equation}
\end{subequations}
where the spin of the quark ($\frac{1}{2}$) and that of the diquark ($s_1 =$ 0 or 1) are coupled to the total spin of the baryon and the same holds for the isospins.

\section{Interacting quark-diquark model of baryons}
In the interacting quark-diquark model of Ref. \cite{Santopinto:2004hw}, the author considers a quark-diquark system, in which the relative motion between the quark and the diquark is described by the relative coordinate $\vec r$ (with conjugate momentum $\vec q$).
In first approximation the relative motion of the quark and the diquark is considered non relativistic, the quark-diquark interaction is assumed to contain a direct and an exchange term, and a mass splitting between the scalar and the axial-vector diquark is also introduced.

The direct interaction contains a Coulomb-like term and a linear confining one:
\begin{equation}
	\label{eqn:direct-inter}
	V_{dir}(r) = - \frac{\tau}{r} + \beta r  \mbox{ } \mbox{ }.
\end{equation}

The author also needs an exchange interaction, since this is a crucial ingredient of a quark diquark model \cite{Lichtenberg:1981pp,Santopinto:2004hw}:
\begin{equation}
	\label{eqn:Vexc}
	V_{ex} = 2A (-1)^{L+1} e^{-\sigma r} \left[ \vec{s}_1 \cdot \vec{s}_2 + \vec{t}_1 \cdot \vec{t}_2 
	+ 2 \vec{s}_1 \cdot \vec{s}_2 \mbox{ } \vec{t}_1 \cdot \vec{t}_2\right]  \mbox{ } \mbox{ }.
\end{equation}
Here $\vec{s}_{1,2}$ and $\vec{t}_{1,2}$ are the spin and isospin operators of the diquark and of the quark, respectively.

The splitting between the two diquark configurations (scalar and axial-vector), called $\Delta$, is parametrized as
\begin{equation}
	\label{eqn:NRsplitting}
	\Delta = B + C \delta_0  \mbox{ } \mbox{ },
\end{equation}
i.e., as a constant $B$, which acts equally in all states with $s_1$ = 1, plus a contact interaction, which only acts on the spatial ground state and of strength $C$.

Summarizing, the Hamiltonian used in this model \cite{Santopinto:2004hw} has the following form:
\begin{equation}
	\label{eqn:HtotNR}
	\left. \begin{array}{rcl}
	\mathcal{H} & = & E_0 + \frac{q^2}{2 \mu} - \frac{\tau}{r} + \beta r + (B + C \delta_0) 
	\delta_{S_{12},1} +  \\
  &  & \left(-1 \right)^{L + 1} 2 A \mbox{ } e^{-\sigma r} \left[ \vec{s}_{1} \cdot \vec{s}_2 \mbox{ } + 
	\vec{t}_1 \cdot \vec{t}_2 + 2 \mbox{ } \vec{s}_1 \cdot \vec{s}_2 \mbox{ } \vec{t}_1 \cdot \vec{t}_2 \right] 
	\mbox{ },  \end{array} \right.  
\end{equation}
where $E_0$ is a constant, $\delta_0$ stands for $\delta_{n0} \delta_{L0}$ (where $n$ is the principal quantum number) and $\mu$ is the reduced mass of the quark-diquark system. The values of the model parameters, fitted to the reproduction of the nonstrange baryon spectrum, are reported in table \ref{tab:Interacting qD model Parameters}.
\begin{table}[h]  
\begin{center}
\begin{tabular}{llllll}
\hline
\hline \\
$\mu$ & $=102$ MeV & $~A$     & $=205$ MeV & $~\sigma$  & =0.60 $fm^{-1}$ \\ 
$\tau$ & $=3.89$   & $~\beta$ & $=0.051~\mbox{fm}^{-2}$ & $~\mathcal{B}$ & $=300$ MeV \\ 
$C$ & $=400$ MeV   & $~E_0$   & $=1706$ MeV & & \\ \\
\hline
\hline
\end{tabular}
\end{center}
\caption{Resulting values for the model parameters.}
\label{tab:Interacting qD model Parameters}
\end{table}  

The total wave functions are combinations of the spin-flavor wave functions with the spatial ones. 
For the central interaction of Eq. (\ref{eqn:direct-inter}), the spatial part of the wave function is
\begin{equation}
	\Psi_{n,L,m}\left(\vec{r}\right) = R_{n,L}(r) Y_{L,m}(\theta,\varphi)  \mbox{ },
\end{equation}
where $R_{n,L}(r)$ is the solution of the radial equation.
For a purely Coulomb-like interaction, such as that of Eq. (\ref{eqn:direct-inter}), the solution of the eigenvalue problem is analytical. All the other interactions are treated as perturbations.
\begin{figure}[htbp]
\begin{center}
\includegraphics[width=9cm]{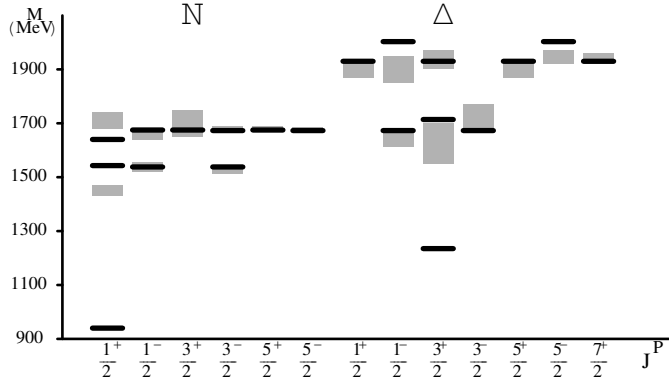}
\end{center}
\caption{\footnotesize{Comparison between the calculated masses \cite{Santopinto:2004hw} (black lines) and the experimental ones (boxes) of non strange $\bigstar \bigstar \bigstar$ and $\bigstar \bigstar \bigstar \bigstar$ baryon resonances \cite{Nakamura:2010zzi}. APS copyright \cite{Santopinto:2004hw}.}}
\label{fig:FIGdiquark}
\end{figure}

In Fig. \ref{fig:FIGdiquark} the results of the model are compared to experimental data, extracted from PDG \cite{Nakamura:2010zzi}.

\section{Relativistic Quark-Diquark Model}
\label{Semirelativistic} 
In Ref. \cite{Ferretti:2011zz} we have performed the relativistic extension of the previous model, within the framework of point form dynamics. The point form dynamics has the interesting property of allowing the coupling of the angular momenta and/or the spins as in the non relativistic case \cite{Klink:1998zz}. The point form formalism and the costruction of the relativistic quark-diquark states are analyzed in details in Refs. \cite{Klink:1998zz,Ferretti:2011zz}.

\subsection{The Mass Operator}
\label{The Model} 
We consider a quark-diquark system, where $\vec{r}$ is the relative coordinate between the quark and the diquark and $\vec{q}$ the conjugate momentum to $\vec{r}$ \cite{Ferretti:2011zz}. 
Our relativistic quark-diquark model is based on the following baryon rest frame mass operator \cite{Ferretti:2011zz}
\begin{equation}
	M = E_0 + \sqrt{q^2 + m_1^2} + \sqrt{q^2 + m_2^2} + M_{\mbox{dir}}(r) + M_{\mbox{cont}}(r) 
	+ M_{\mbox{ex}}(r) ~,  
	\label{eqn:H0}
\end{equation}
where $E_0$ is a constant, $M_{\mbox{dir}}(r)$ and $M_{\mbox{ex}}(r)$ are the direct and the exchange diquark-quark interaction respectively, $m_1$ and $m_2$ are the diquark and quark masses ($m_1$ is either $m_S$ or $m_{AV}$ according if the mass operator acts on a scalar or an axial-vector diquark) and $M_{\mbox{cont}}(r)$ is a contact interaction.

The direct term is the sum of a Coulomb-like interaction with a cut-off and a linear confinement term:
\begin{equation}
  \label{eqn:Vdir}
  M_{\mbox{dir}}(r)=-\frac{\tau}{r} \left(1 - e^{-\mu r}\right)+ \beta r ~~.
\end{equation}

As in Ref. \cite{Santopinto:2004hw}, we also need an exchange interaction, since this is a crucial ingredient of a quark diquark model \cite{Lichtenberg:1981pp,Santopinto:2004hw}:
\begin{equation}
  M_{\mbox{ex}}(r) =  (-1)^{l + 1} e^{-\sigma r} 
  \left [ A_S \mbox{ } \vec s_1 \cdot \vec s_2 + A_I \mbox{ } \vec t_1 \cdot \vec t_2 
  + A_{SI} \mbox{ } \vec s_1 \cdot \vec s_2 \mbox{ } \vec t_1 \cdot \vec t_2 \right ]  ~~.
    \label{eqn:VexchS1}
\end{equation}
Here $\vec{s}_i$ and $\vec{t}_i$ ($i=1,2$) are the spin and isospin operators.

Furthermore, we consider a contact interaction similar to that introduced by Godfrey and Isgur in Ref. \cite{Godfrey:1985xj}, i.e.
\begin{equation}
	\label{eqn:Vcont(r)}
	M_{\mbox{cont}} = \left(\frac{m_1 m_2}{E_1 E_2}\right)^{1/2+\epsilon}  
	\frac{\eta^3 D}{\pi^{3/2}} e^{-\eta^2 r^2} \mbox{ } \delta_{L,0} \delta_{s_1,1} 
	\left(\frac{m_1 m_2}{E_1 E_2}\right)^{1/2+\epsilon} \mbox{ },
\end{equation}
where $E_i = \sqrt{q^2 + m_i^2}$ ($i$ = 1, 2), $\epsilon$, $\eta$ and $D$ are parameters of the model \cite{Ferretti:2011zz}.

The whole mass operator of Eq. (\ref{eqn:H0}) has been diagonalized by means of a numerical variational procedure, based on harmonic oscillator trial wave functions. With a variational basis made of $N = 256$ harmonic oscillator shells the results converge very well, even if we have noticed that convergence is already satisfying for $N \approx 150$.
Finally, it has to be noted that in Ref. \cite{Ferretti:2011zz} all the calculations are performed without any perturbative approximation.
 

\subsection{Nonstrange baryon spectrum}
\label{Results and discussion}
\begin{figure}[htbp]
\begin{center}
\includegraphics[width=9cm]{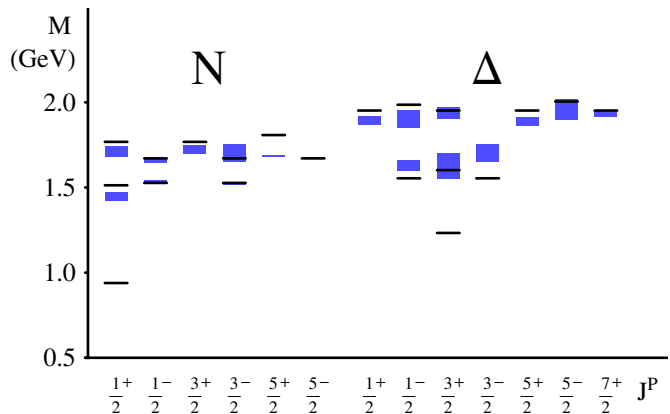}
\end{center}
\caption{Comparison between the calculated masses \cite{Ferretti:2011zz} (black lines) of the $3\bigstar$ and $4\bigstar$ non strange baryon resonances (up to 2 GeV) and the experimental masses from PDG \cite{Nakamura:2010zzi} (boxes). APS copyright \cite{Ferretti:2011zz}.}
\label{fig:Spectrum3e4}
\end{figure}

Fig. \ref{fig:Spectrum3e4} shows our results for the non strange baryon spectrum, obtained with the set of parameters of table \ref{tab:ResultingParameters} \cite{Ferretti:2011zz}. One can notice that the overall quality in the reproduction of the experimental data ($3 \bigstar$ and $4 \bigstar$ resonances) is comparable to that of other relativistic constituent quark models \cite{Capstick:1986bm,Loring:2001kx}, even if our results are not plagued with the problem of missing resonances. Indeed our model does not predict missing states below the energy of 2 GeV \cite{Ferretti:2011zz}.
\begin{table}[h]  
\begin{center}
\begin{tabular}{llllll}
\hline
\hline \\
$m_q$ & $=200$ MeV & $~m_{S}$ & $=600$ Mev & $~m_{AV}$ & $=950$ MeV \\ 
$~\tau$ & $=1.25$   & $~\mu$ & $=75.0~\mbox{fm}^{-1}$ & $~\beta$ & $=2.15~\mbox{fm}^{-2}$ \\ 
$A_S$ & $=375$ MeV & $A_I$ & $=260$ MeV & $A_{SI}$ & $=375$ MeV \\
$~\sigma$ & $=1.71~\mbox{fm}^{-1}$ & $~E_0$ & $=154$ MeV & $D$ & $=4.66$ $fm^2$ \\ 
$~\eta$ & $=10.0~\mbox{fm}^{-1}$ & $~~\epsilon$ & $=0.200$ &  &  \\ \\
\hline
\hline
\end{tabular}
\end{center}
\caption{Resulting values for the model parameters \cite{Ferretti:2011zz}.}
\label{tab:ResultingParameters}
\end{table}  

An important aspect in a quark-diquark model is the mass difference between axial-vector and scalar diquarks. Indeed, while the values of the diquark masses are model dependent, their difference is not. 
It is thus worthwhile noting that our estimation \cite{Ferretti:2011zz} lies within the mass range predicted by the majority of the studies on the quark-diquark model and diquark correlations \cite{Wilczek:2004im,Jaffe:2004ph,Orginos:2005vr,Babich:2007ah,Schafer:1993ra,Burden:1996nh,Cahill:1995ka,Flambaum:2005kc,Bloch:1999ke,Eichmann:2008ef,Lichtenberg:1979de,deCastro:1993sr}.
Such evaluations come from phenomenological observations \cite{Wilczek:2004im,Jaffe:2004ph}, lattice QCD calculations \cite{Orginos:2005vr,Babich:2007ah}, instanton liquid model calculations \cite{Schafer:1993ra}, applications of Dyson-Schwinger, Bethe-Salpeter and Fadde'ev equations \cite{Burden:1996nh,Cahill:1995ka,Flambaum:2005kc,Bloch:1999ke,Eichmann:2008ef} and constituent quark-diquark model calculations \cite{Lichtenberg:1979de,deCastro:1993sr}.

\section{Nucleon electromagnetic form factors}
The nucleon elastic electromagnetic (e.m.) form factors contain important informations on the internal structure of the nucleon and thus can be a useful tool for the understanding of the strong interaction. This is the reason why an extensive program for their experimental determination is currently underway at several facilities around the world. 

A renewed interest in the description of the internal structure of the nucleon has been triggered by the most recent results on the ratio of the electric and magnetic form factors of the proton \cite{Milbrath:1997de,Jones:1999,Dieterich:2000mu,Pospischil:2001pp,Gayou:2001qt,Gayou:2001qd,Punjabi:2005wq,Puckett:2010ac,Meziane:2010xc,Puckett:2011xg}, showing an unexpected decrease with $Q^2$, at variance with the widely accepted dipole-fit expectation. 
The observed deviation of the ratio $\mu_p G_E^p(Q^2)/G_M^p (Q^2)$ from the dipole-fit expectation, i.e. $\mu_p G_E^p(Q^2)/G_M^p (Q^2) = 1$, can be expected in CQM's provided that relativistic effects are taken into account.

In light of this we have decided to calculate the nucleon elastic electromagnetic form factors in the relativistic quark-diquark model \cite{Ferretti:2011zz}. 
The e.m. form factors have been computed in the point form spectator impulse approximation (PFSA) \cite{Klink:1998qf}, using as nucleon state that obtained by solving the eigenvalue problem of the mass operator (\ref{eqn:H0}) through a numerical variational procedure \cite{Ferretti:2011zz}. 
The diquark is considered as a particle with spin $s_1=0$ (and isospin $t_1=0$), being a scalar one.
There are also calculations of the nucleon e.m. form factors within point form dynamics for the three constituent quark case. For example see Ref. \cite{DeSanctis:2004ej,Sanctis:2007zz,Santopinto:2010zz}, where the nucleon e.m. form factors have been calculated with a relativized version of the hypercentral quark model.
 
In the impulse approximation the electromagnetic properties of composite particles should be determined by the e.m. properties of their constituents. Thus the e.m. properties of hadrons should be determined by the e.m. properties of constituent quarks (and diquarks) \cite{Klink:1998qf}: this means approximating the e.m. current operator $J^{\mu}(0)$ by one body operators, assuming that the matrix elements of the many body current operator are smaller than the one body ones. 

For the calculation we choose the Breit-frame where the initial and final nucleon momenta are taken along the $z$-axis, that is $\vec P_N=\vec P'_N=-\frac{q}{2}\hat z$, being $q$ the $z$-component of the virtual photon momentum.
The PFSA matrix elements of the e.m. current operator, in the space of the single particle free states, take the form \cite{DeSanctis:2011zz}
\begin{equation}
	\label{eq:fcme}
	\begin{array}{l}
		\left\langle p'_1,p'_2,\lambda'_1,\lambda'_2 \right| J^\mu \left| p_1,p_2,\lambda_1,\lambda_2 \right\rangle = 
		\bar u_{\lambda'_2}(p'_2) u_{\lambda_2}(p_2) j^\mu_1(p'_1,p_1) \delta^3(p'_2-p_2) \\
		\hspace{1cm} + \mbox{ } \bar u_{\lambda'_2}(p'_2) j^\mu_2 u_{\lambda_2}(p_2) \delta(p'_1-p_1)  
		\mbox{ },
	\end{array}
\end{equation}
where the first and the second term represent the diquark current (acting the quark as a spectator) and the quark current (acting the diquark as a spectator), respectively.

Both the diquark and the quark are considered as effective, composite, constituent particles, so that form factors are added to their vertex functions.
In the PFSA, the interacting particle is considered on-shell, so that, for the $s_1=0$ diquark current [$j^\mu_1(p'_1,p_1)$], we need one form factor, $F_1^D(\tilde q^2)$, where $\tilde q^2$ is the squared four-momentum effectively exchanged by the interacting constituents. 
On the other hand, Dirac and Pauli form factors $F_1^q(\tilde q^2)$ and $F_2^q(\tilde q^2)$ are used to parametrize the quark current [$j^\mu_2(p_2',p_2)$] \cite{DeSanctis:2011zz}.

We enforce phenomenologically current conservation by introducing a conserved current operator $J_\mu^c$, defined as
\begin{equation}
	\label{eqn:conserved-current-op.}
	J_\mu^c = J_\mu + \frac{q_\mu \mbox{ } (q \cdot J)}{Q^2}  \mbox{ },
\end{equation}
where $Q^2 = -q^2$.
It is worthwhile noting that, for elastic scattering in the Breit frame, the modified current operator of Eq. (\ref{eqn:conserved-current-op.}) does not change the results for the e.m. form factors of the nucleon.

We compute the matrix elements of the nucleon current following the standard formalism of the PFSA \cite{Klink:1998zz}. The invariant nucleon form factors are given by \cite{DeSanctis:2011zz}
\begin{equation}
	\label{eq:fff}
	\begin{array}{l}	
		G^\nu_{\mu' \mu} = \int d^3k_2' d^3k_2 \mbox{ } \psi^*(\vec k'; \mu') \psi(\vec k; \mu) 
		\left\langle k'_1 \right| j_1^\nu \left| k_1 \right\rangle 
		D^{1/2}_{\mu' \mu}[R_W(k_2, B^{-1}(v_f) B(v_i))] \delta^3(k'_2 -B^{-1}(v_f) B(v_i)k_2) \\
		\hspace{0.5cm} + \mbox{ } \int d^3k_1' d^3k_1 \mbox{ } \psi^*(\vec k'; \mu') \psi(\vec k; \mu) 
		D^{1/2*}_{\lambda'_2 \mu'}[R_W(k'_2, B(v_f))] \left\langle k'_2,\lambda'_2\right| j_2^\nu 
		\left|k_2,\lambda_2 \right\rangle D^{1/2}_{\lambda_2 \mu}[R_W(k_2, B(v_i))]  \\
		\hspace{0.5cm} \times \mbox{ } \delta^3(k'_1-B^{-1}(v_f) B(v_i)k_1)   \mbox{ },
	\end{array}
\end{equation}
where the first term represents the diquark contribution and the second term the quark one, $\psi(\vec k; \mu)$ is the quark-diquark system wave function, with the overall velocity factored out and $\mu$ is the third component of the nucleon spin.
\begin{figure*}[!ht]
  \includegraphics[width=8cm]{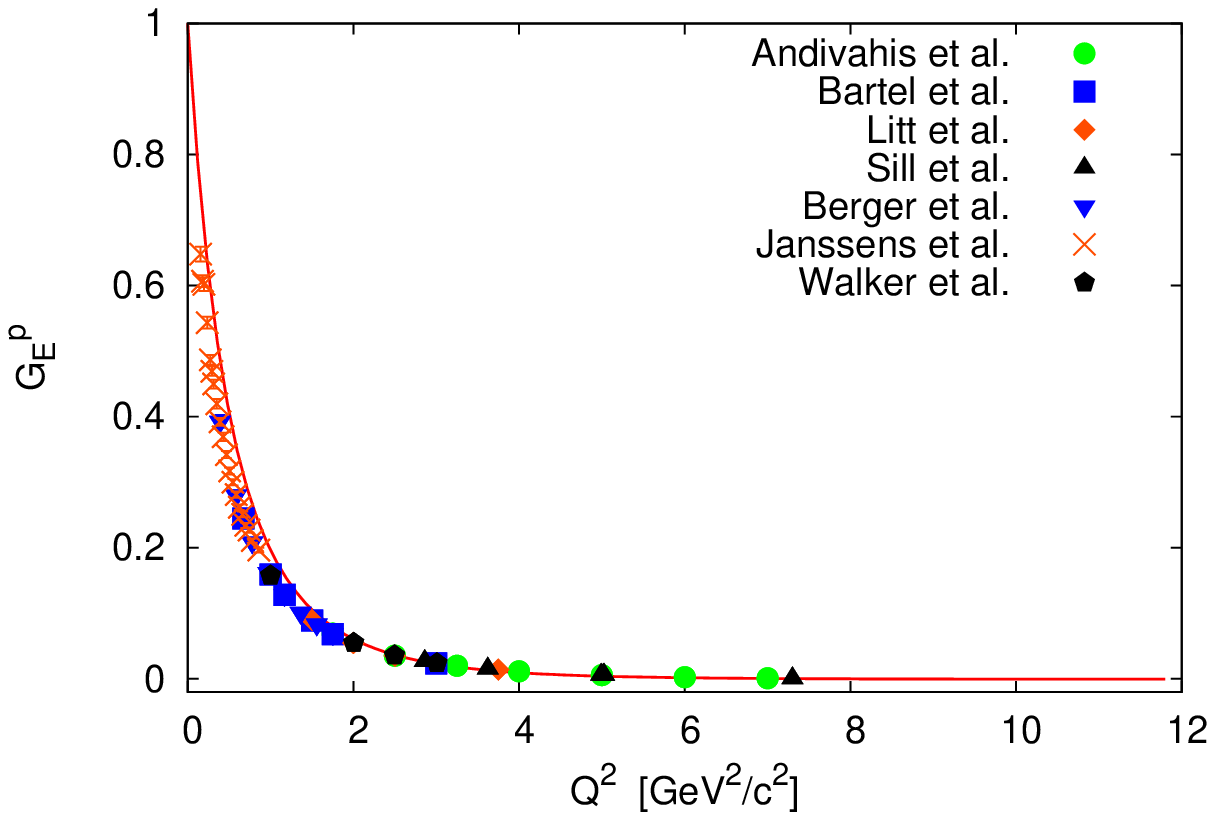} 
  \includegraphics[width=8cm]{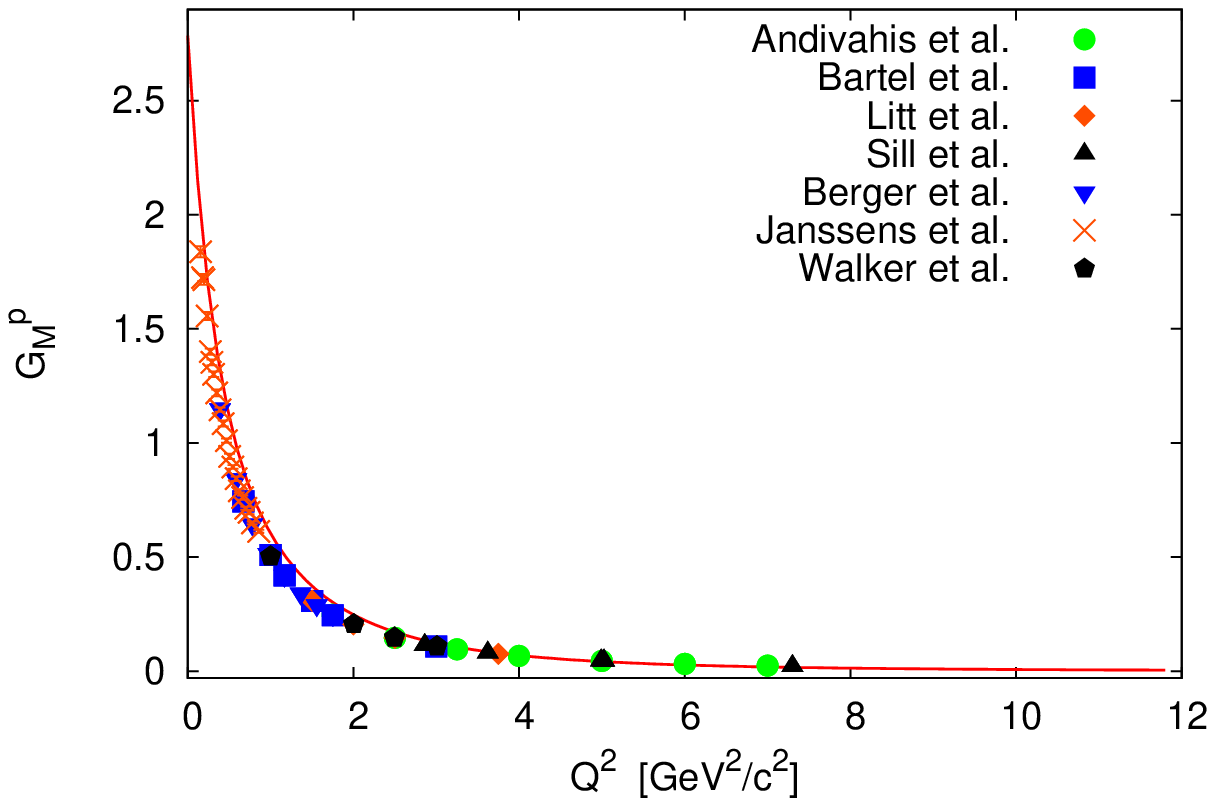} 
  \includegraphics[width=8cm]{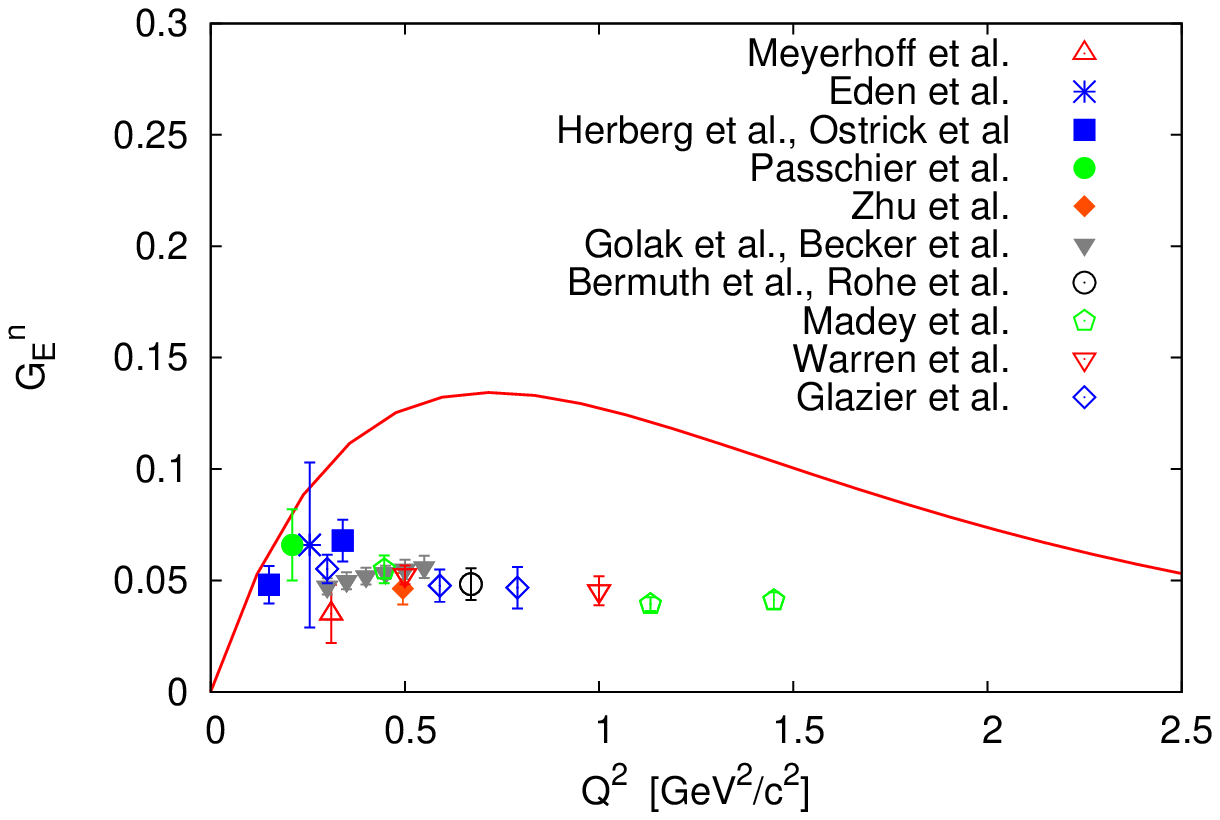} 
  \includegraphics[width=8cm]{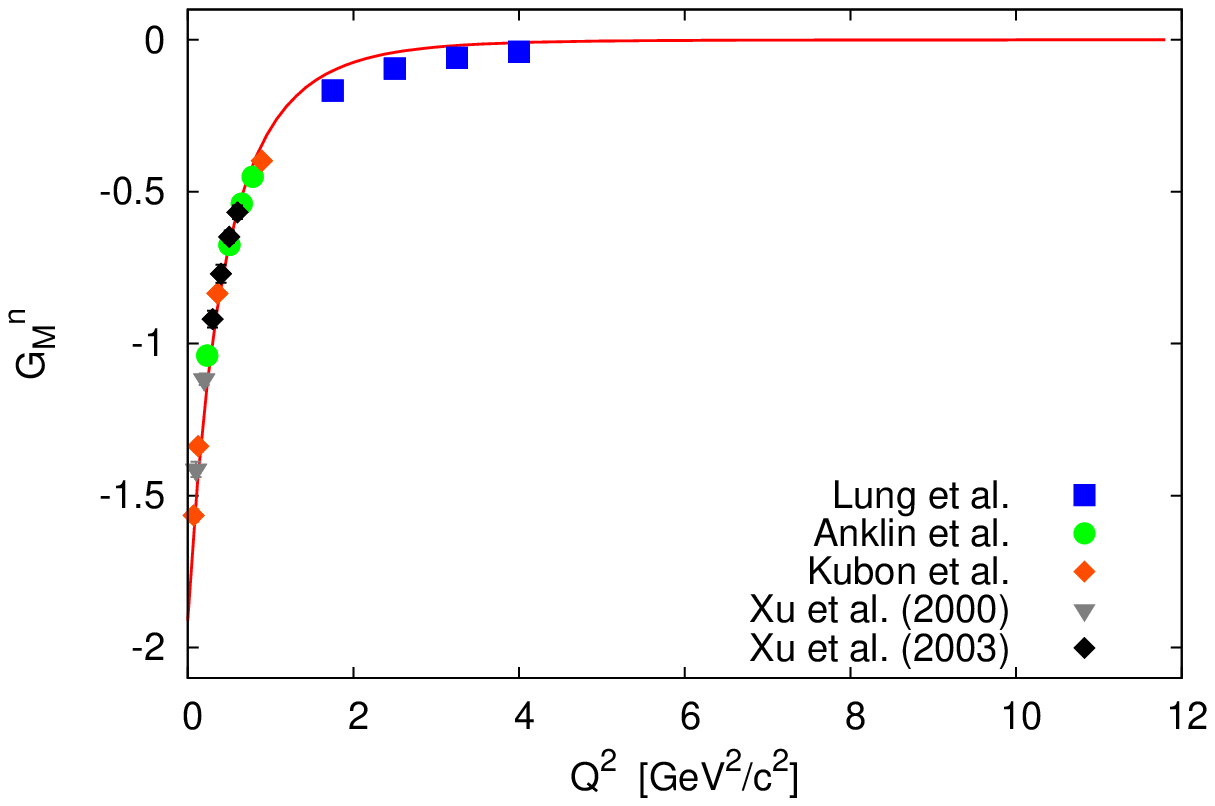}
\caption{Elastic form factors of the nucleon. The solid line corresponds to the relativistic quark-diquark calculation of Ref. \cite{DeSanctis:2011zz}. The experimental data for $G_M^p$ are taken from the reanalysis made by Brash {\em et al.} \cite{Brash:2004} of the data from Refs. \cite{Andivahis:1994,Bartel:1973,Janssens:1966,Litt:1970,Sill:1993,Berger:1971,Walker:1994}; the points shown for $G_E^p$ are obtained from the data on $G_M^p$ and the linear fit \cite{Brash:2004} of the Jlab data on the ratio $\mu_pG_E^p/G_M^p$; for $G_E^n$ the experimental data are taken from Refs. \cite{Meyerhoff:1994,Eden:1994,Herberg:1999,Passchier:1999,Zhu:2001,Golak:2001,Bermuth:2003,Madey:2003,Warren:2004,Glazier:2005} and for $G_M^n$ from Refs. \cite{Anklin:1994ae,Kubon:2002,Lung:1993,Markowitz:1993,Rock:1982,Xu:2000,Xu:2003}. APS copyright \cite{DeSanctis:2011zz}.}
\label{fig:ff_nudo_mds}
\end{figure*}
The $\tilde q^2$ behavior of the quark and diquark form factors is chosen as \cite{DeSanctis:2011zz}
\begin{equation}
	F_D(\tilde q^2) = \frac{1}{3} \frac{1}{\left(1-\frac{\tilde q^2}{A_D^2}\right)^{B_D}}  \mbox{ },
	\hspace{0.5cm}
	F_1^{u,d}(\tilde q^2) = \frac{e_{u,d}}{\left(1-\frac{\tilde q^2}{A_{u,d}^2}\right)^{B_{u,d}}}  \mbox{ },
	\hspace{0.5cm}
	F_2^{u,d}(\tilde q^2) = \frac{\kappa_{u,d}}{\left(1-\frac{\tilde q^2}{C_{u,d}^2}\right)^{D_{u,d}}}  \mbox{ },
\end{equation}
where $e_{u,d}$ and $\kappa_{u,d}$ are the electric charge and the anomalous magnetic moment of the $u/d$ quarks, respectively.

\begin{figure}[htbp]
\begin{center}
\includegraphics[width=8cm]{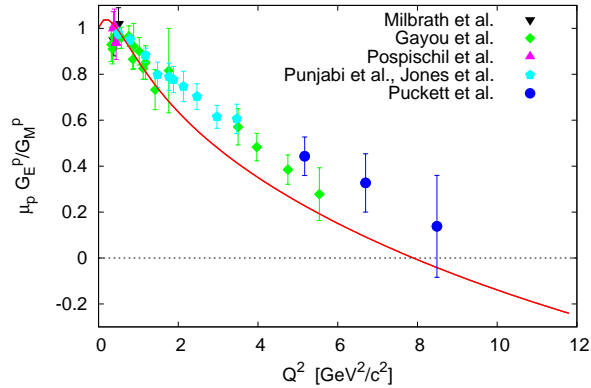}
\end{center}
\caption{Ratio $\mu_p G_E^p(Q^2)/G_M^p (Q^2)$. The experimental data are taken from Ref. \cite{Jones:1999,Gayou:2001qt,Gayou:2001qd,Milbrath:1997de,Pospischil:2001pp}. The solid line corresponds to the relativistic quark-diquark calculation of Ref. \cite{DeSanctis:2011zz}. APS copyright \cite{DeSanctis:2011zz}.}
\label{fig:rap}
\end{figure}

By fitting the free parameters to the reproduction of $G_M^p$, $G_M^n$, $G_E^n$ and the ratio $\mu_p G_E^p/G_M^p$ we obtain the curves shown in Figs. \ref{fig:ff_nudo_mds} and \ref{fig:rap} (see also table \ref{tab:FFs-Parameters} for the values of the free parameters).
\begin{table}[h]  
\begin{center}
\begin{tabular}{llllll}
\hline
\hline \\
$A_D$ & = 1.27 GeV & $~B_D$ & = 2.01 &             &          \\ 
$A_u$ & = 1.68 GeV & $~B_u$ & = 0.66 &             &          \\ 
$A_d$ & = 1.50 GeV & $~B_d$ & = 1.40 &             &          \\
$C_u$ & = 0.72 GeV & $~D_u$ & = 1.34 & $~\kappa_u$ & = 0.38   \\
$C_d$ & = 1.20 GeV & $~D_d$ & = 1.95 & $~\kappa_d$ & = -0.33  \\ \\
\hline
\hline
\end{tabular}
\end{center}
\caption{Resulting values for the free parameters in quark and diquark form factors \cite{DeSanctis:2011zz}.}
\label{tab:FFs-Parameters}
\end{table}  

As it can be seen in Figs. \ref{fig:ff_nudo_mds} and \ref{fig:rap}, the experimental data are reasonably well reproduced. 
The discrepancies arising in the case of the electric neutron form factor are related to the way in which the SU(6) symmetry is broken and this represents an open problem of CQM's. 
With respect to the non relativistic case, the relativistic wave functions have more high momentum components.
This fact, together with the application of exact boosts to the Breit-Frame, leads to a reasonable reproduction of the existing data on the electromagnetic form factors. 
To obtain a good reproduction of the experimental data, it has also been necessary to develope the electromagnetic current formulas directly with diquark and quark form factors, to take the effective nature of the constituent quark and diquark into account.
The introduction of such phenomenological form factors in the electromagnetic current makes it possible to obtain a reasonable agreement with the available experimental data up to 8 GeV$^2$ and a crossing of the zero for the ratio $\mu_p G_E^p/G_M^p$ at $Q^2 \approx 8$ GeV$^2$. 

\end{document}